\documentclass[a4paper, 11pt]{article}

\def\b{\begin{eqnarray}}
\def\e{\end{eqnarray}}
\def\n{\noindent}
\begin{document}


\begin{center}

{\huge \textbf{Reissner--Nordstr\" om Expansion \\}}
\vspace {10mm} \noindent {\large \bf
Emil M. Prodanov$^{\, \, \spadesuit \, , \, \clubsuit \, , \, \heartsuit \, , \, \, \ast}$,
Rossen I. Ivanov$^{\, \, \spadesuit \, , \, \diamondsuit \, , \, \, \dagger}$,  \vskip.5cm
and V.G. Gueorguiev$^{\, \, \Delta \, , \, \diamondsuit \, , \, \, \ddagger}$}

\vskip1cm \n
\noindent
\begin{tabular}{c}
$\phantom{e}^\spadesuit ${\it \,\, School of Mathematics, Trinity College, University of Dublin, Ireland} \\
\\
$\phantom{e}^\clubsuit ${\it \,\, School of Mathematical Sciences, Dublin Institute of Technology, Ireland} \\
\\
$\phantom{e}^\heartsuit ${\it \,\, School of Theoretical Physics, Dublin Institute of Advanced Studies, Ireland} \\
\\
$\phantom{R^R}^\Delta ${\it \,\, Lawrence Livermore National Laboratory, Livermore, CA 94550, USA} \\
\\
$\phantom{e}^\diamondsuit${\it \,\, On Leave of Absence from Institute for Nuclear Research and Nuclear Energy,} \\
$\phantom{e^\diamondsuit}${\it \,Bulgarian Academy of Sciences, 72 Tzarigradsko Chaussee, Sofia--1784, Bulgaria} \\
\\
$\phantom{.}^\ast$ {\it prodanov@maths.tcd.ie} \\
$\phantom{.}^\dagger$ {\it ivanovr@tcd.ie} \\
$\phantom{.}^\ddagger$ {\it vesselin@mailaps.org} \\
\end{tabular}
\vskip1cm
\end{center}

\begin{abstract}
\n
We propose a classical mechanism for the cosmic expansion during the radiation-dominated era. This mechanism assumes
that the Universe is a two-component gas. The first component is a gas of ultra-relativistic "normal" particles
described by an equation of state of an ideal quantum gas of massless particles. The second component consist of
"unusual" charged particles (namely, either with ultra-high charge or with ultra-high mass) that provide the
important mechanism of expansion due to their interaction with the "normal" component of the gas. This interaction
is described by the Reissner--Nordstr\" om metric purely geometrically --- the ``unusual'' particles are modeled as
zero-dimensional naked singularities inside spheres of gravitational repulsion. The radius of a repulsive
sphere is inversely proportional to the energy of an incoming particle or the temperature. The expansion
mechanism is based on the inflating of the "unusual" particles (of charge $Q$) with the drop of the
temperature --- this drives apart all neutral particles and particles of specific charge $q/m$ such that
$\mbox{sign}(Q) q/m  \ge - 1$. The Reissner--Nordstr\" om expansion naturally ends at recombination. We discuss the
range of model parameters within which the proposed expansion mechanism is consistent with the restrictions regarding
quantum effects.

\end{abstract}

\newpage

\n
We propose a classical mechanism for the expansion of the Universe during the radiation-dominated era. Various
mechanisms have been proposed for the cosmic expansion. It is generally accepted that a scalar field drove the
inflation of the Universe (see, for example \cite{guth} and the references therein) and that the expansion during
the inflation was exponential or power law \cite{lucchin}. \\
Brisudova {\it et al.} \cite{brisudova} considered a cosmological model with a complex scalar field, minimally coupled
to a U(1) gauge field. The expansion of the Universe was generated by a long-range repulsive force resulting from
the endowment of the photon with a mass depending on the scalar field. \\
Our model is based on the assumption that the expanding Universe is a two-component gas. The first component is a gas
of ultra-relativistic "normal" particles described by an equation of state of an ideal quantum gas of massless particles.
We show that the expansion of the Universe could be due to the second fraction of the gas --- a component consisting
of "unusual" charged particles (namely, either with ultra-high charge or with ultra-high mass) that provide the
important mechanism of expansion due to their interaction with the "normal" component of the gas. The "unusual" particles
are naked singularities and the interaction mechanism is the gravitational repulsion of the naked singularities. Naked
singularities are particles of charge $Q$ greater than their mass $M$ (in geometrized units $G = 1 = c$) and we
model them as zero-dimensional Reissner--Nordstr\" om \cite{rn, mtw} gravitational entities surrounded by spheres of
gravitational repulsion \cite{cohen}. In our picture, the Universe has local Reissner--Nordstr\" om geometry, but
globally, the geometry is that of Robertson--Walker \cite{mtw, rw}. Namely, we confine our attention to the local
spherical neighbourhood of a single naked singularity and consider the Universe as multiple copies (fluid) of such
neighbourhoods. We show that the radii of the repulsive spheres grow in inverse proportion with the temperature --- the
"unusual" particles ``grow'' as the temperature drops and drive away the "normal" fraction of the Universe. This
repulsion results in power law expansion with scale factor $a(\tau) \sim \sqrt{\tau}$, corresponding to the expansion
during the radiation-dominated era. The gravitational repulsion is not powerful enough to achieve accelerated expansion
that solves the horizon problem and thus account for the inflation the Universe [$a(\tau) \sim e^{H\tau}$ or
$a(\tau) \sim \tau^n$, with $n > 1$], unless charge non-conservation is involved. \\
We determine a particle's ``radius'' by calculating the turning radius of a radially moving incoming (charged) test
particle of ultra-high energy $kT \gg mc^2$ (where $m$ is the test particle's rest mass). The proposed model is
simplified significantly by considering the incoming particles as collisionless probes rather than involving
their own gravitational fields and by not considering the more general and physically more relevant Kerr--Newman
geometry \cite{mtw, kerr, newman}, thus ignoring magnetic effects caused by rotation of the centre, which drags the
inertial reference frames, and the particles' spins. \\
In 1971, Hawking suggested \cite{hawk} that a large number of gravitationally collapsed charged objects of very low mass
(of the order of Planck's mass) were formed in the early Universe. Hawking argues that gravitational collapse is,
essentially, a classical process and a black hole cannot form if its Schwarzschild radius is smaller than the Planck
length $(G h c^{-3})^{1/2} \sim 10^{-35}$ m (or its mass --- smaller than $10^{-8}$ kg), since, at Planck lengths,
quantum fluctuations of the metric become relevant. However, for lengths larger than $10^{-35}$ m, it is legitimate
to ignore quantum gravitational effects and treat the metric classically. One would expect that a collapsed object
could form if the Schwarzschild radius is greater than the Compton wavelength $h/(mc)$ of one of the elementary
particles which formed it. This corresponds to a minimum mass of $10^{11}$ kg. However, Hawking argues further, this
is not the case, since the Compton wavelength of a photon is infinite, yet a sufficient concentration of
electromagnetic radiation can cause gravitational collapse. Hawking suggests that the wavelength to be considered
should not be the wavelength at rest, but $hc/E$, where $E$ is the typical energy of the particle. For
ultra-relativistic particles, $E \sim kT \gg mc^2$. \\
We will show that the "radius" of the "unusual" particles is inversely proportional to the temperature of the Universe:
$r_0(T) = Q (q + m)/(kT)$ (where $Q$ is the charge of the ``unusual'' particles, $m$ and $q$ are the mass and charge
of the ``normal'' particles respectively). This is the characteristic length that is to be considered, instead of a
Schwarzschild radius, and compared to the wavelength  $\lambda(T) \sim hc/(kT)$. Our classical model is applicable only
when quantum gravitational effects do not take place between the ``unusual'' particles and the ``normal'' particles.
Therefore, the range of validity of the model is determined by the condition that the ``radius'' $r_0(T)$ of an
``unusual'' particle exceeds the wavelength $\lambda(T)$. Thus, for a ``normal'' particle of typical charge $e$ (the
electron's charge), the repulsive centre must have charge $Q$ higher than $h/e$ (in geometrized units). If quantum
gravitational effects between the ``unusual'' particles and neutral ``normal'' particles are to be negligible, the
repulsive centre must have charge $Q$ higher than $h/m$ (again, in geometrized units). It is, understandably, hard
to comprehend the conditions under which particles of such ultra-high charge could have formed since, for a gravitational
collapse, huge gravitational energy will be needed to overcome the electrical repulsion. The formation of particles of
such ultra-high charge is an open issue for the model. \\
Fortunately, there is an alternative to particles of ultra-high charge. These will be particles of ultra-high mass and
charge which is comparable to the electron charge $e$. Of course, the mass of such particle must not exceed its charge
so that a naked singularity treatment exists. This puts an upper limit on the mass $M$ of these particles at
$10^{21}$ electron masses or $10^{-9}$ kg. As the rest mass of these particles is so huge and comparable to $kT$ for
quite high temperature ($\sim 10^{31}\!$ K), the characteristic length that is to be considered for quantum effects is
$\lambda \sim h/M$ (or $h/e$, as $M$ is of the order of $e$, without exceeding it). One can immediately determine the
temperature below which the expansion mechanism with "unusual" particles of mass $M \sim 10^{-9}$ kg is valid:
$\, T \, < \, e^3/(hk) \, \sim 10^{29}\!$ K [for wavelength $\lambda \sim h/e$ not larger than the particle's "radius"
$r_0(T) \sim e^2/(kT)$]. \\
Since the early Universe was very dense and ultra-relativistic, we may speculate that not two-body, but many-body
collisions of "normal" particles led to the production of particles with $Q > M$. This is hardly a quantitative
explanation, however, the existence of such ultra-heavy particles has been studied by many authors. Of particular
importance is the work of de Rujula, Glashow and Sarid \cite{glashow}. The authors consider that very heavy charged
particles (CHAMPS), which have survived annihilation, were produced in the early Universe. These CHAMPS are even viewed
as dark matter candidates. Secondly, Preskill has shown \cite{preskill} that ultra-heavy magnetic monopoles were created
so copiously in the early Universe that they outweighed everything else in the Universe by a factor of $10^{12}$.
Again, these ultra-heavy monopoles can serve as the "unusual" fraction of the two-component gas. This time $Q$
will be the magnetic charge. The other fraction of the gas will then consist of magnetically neutral particles. \\
Not all particles are repelled by naked singularities. If the specific charge $q/m$ of a probe is such that
$\mbox{sign}(Q) q/m  < - 1$, the probe will reach the singularity \cite{cohen}. In result, the absolute value of the
charge $Q$ of the naked singularity will diminish, while the mass $M$ of the naked singularity will increase. If this
process is repeated a sufficient number of times, the naked singularity will annihilate --- it will pick up a horizon
and turn into a black hole. We assume that the "unusual" particles have survived such annihilation or have annihilated
at a very slow rate. Thus, as the "radius" of an "unusual" particle grows in inverse proportion with the temperature,
the Reissner--Nordstr\" om expansion would continue forever. This is not the case however --- the Reissner--Nordstr\"om
expansion ends naturally at recombination. For a neutral "normal" particle, at a certain distance from the "unusual"
particle, the gravitational repulsion turns into gravitational attraction, while the gravitational repulsion of a charged
"normal" particle extends to infinity (the gravitational attraction is not sufficiently strong to overcome the electrical
repulsion) \cite{cohen}. At recombination, charged particles --- which have so far been being repelled by the naked
singularities --- combine to form neutral particles. These neutral particles are formed sufficiently far from the naked
singularities --- beyond the region of gravitational repulsion --- and this stops the Reissner--Nordstr\" om expansion
mechanism.

\vskip.5cm
\n
We consider the general motion of a particle in Kerr--Newman geometry \cite{kerr, newman, mtw}. The Kerr--Newman
metric in Boyer--Lindquist coordinates \cite{bl} and geometrized units is given by:
\b
ds^2 & = & - \, \, \frac{\Delta}{\rho^2} (dt - a \sin^2 \theta \,\, d\phi)^2 +
\frac{\sin^2 \theta}{\rho^2} \Bigl[ a \, dt  - (r^2 + a^2) d\phi \Bigr]^2 \nonumber \\
& & \hskip6cm + \, \, \frac{\rho^2}{\Delta} dr^2 + \rho^2 d \theta^2 \, ,
\e
where
\b
\Delta & = & r^2 - 2 M r + a^2 + Q^2\, , \\
\rho^2 & = & r^2 + a^2 \cos^2 \theta \, .
\e
In the above, $M$ is the mass of the centre, $a > 0$ --- the specific angular momentum of the centre
(i.e. angular momentum per unit mass) and $Q$ --- the charge of the centre. \\
The motion of a particle of mass $m$ and charge $q$ in gravitational and electromagnetic fields is governed
by the Lagrangian:
\b
L = \frac{1}{2} \, \, g_{ij} \frac{d x^i}{d \lambda} \frac{d x^j}{d \lambda} - \frac{q}{m} \, A_i
\, \frac{dx^i}{d\lambda} \, .
\e
In the above, $\lambda$ is the proper time $\tau$ per unit mass $m$: $\lambda = \tau/m$ and $A$ is the
vector electromagnetic potential, determined by the charge $Q$ and specific angular momentum $a$ of the centre:
\b
A_i dx^i = - \frac{Q r}{\rho^2} (dt - a \, \sin^2 \theta \, d \phi) \, .
\e
(The magnetic field is due to the dragging of the inertial reference frames into rotation around the centre.) \\
The equations of motion for the particle are:
\b
\label{geo}
\frac{d^2 x^i}{d \tau^2} + \Gamma^i_{jk} \, \frac{dx^j}{d \tau} \, \frac{dx^k}{d \tau} =
\frac{q}{m} F^i_{\phantom{ii}j} \, \frac{d x^j}{d \tau} \, ,
\e
where $F = dA$ is Maxwell's electromagnetic tensor and $\Gamma^i_{jk}$ are the Christoffel symbols. \\
For Kerr--Newman geometry, the geodesic equations (\ref{geo}) can be written as \cite{car} (see also \cite{fn}):
\b
\rho^2 \, \frac{dt}{d \lambda} &  =  & - a^2 E \sin^2 \theta + a J
    + \frac{r^2 + a^2}{\Delta} \Bigl[ E(r^2 + a^2) - J a - q Q r \Bigr] \, ,  \\ \nonumber \\
\label{potential}
\rho^2 \, \frac{dr}{d \lambda} & = & \pm \sqrt{\Bigl[ E(r^2 + a^2) - J a - q Q r \Bigr]^2
    - \Delta \Bigl[ m^2 r^2 + (J - a E)^2 + K \Bigr]} \, , \nonumber \\ \\
\rho^2 \, \frac{d\theta}{d \lambda} & =  & \pm \sqrt{K - \cos^2 \theta \Bigl[ a^2(m^2 - E^2)
    + \frac{1}{\sin^2 \theta} \, J^2 \Bigr]} \, , \\ \nonumber \\
\rho^2 \, \frac{d\phi}{d \lambda} &  =  & - a E + \frac{J}{\sin^2 \theta}
    + \frac{a}{\Delta} \Bigl[ E(r^2 + a^2) - J a - q Q r \Bigr] \, ,
\e
where $\, E = (1/m) \partial L / \partial \dot{t} \, \, $ is the conserved energy of the particle,
$J = (1/m) \partial L / \partial \dot{\phi} \, \, $ is the conserved projection of the particle's angular
momentum on the axis of the centre's rotation (dots denote differentiation with respect to $\tau$).  $K$ is another
conserved quantity given by:
\b
K = p_\theta^2 + \cos^2 \theta \Bigl[ a^2(m^2 - E^2)+ \frac{1}{\sin^2 \theta} \, J^2 \Bigr].
\e
Here $p_\theta = (1/m) \partial L / \partial\dot{\theta}$ is the $\, \theta$-component of the particle's
four-momentum.  \\
\n
For simplicity, we will confine our attention to the radial motion of a charged test particle in
Reissner--Nordstr\" om \cite{rn} geometry  (see \cite{cohen} for a very thorough analysis). In other words,
we will request $\dot{\theta} = 0 = \dot{\phi}$ and also set $a = 0$. \\
Equation (\ref{potential}) then reduces to:
\b
\label{eq}
\dot{r}^2 =  r^{-2} \Bigl[(\epsilon^2 - 1) r^2 + 2 \Bigl(1 - \frac{q}{m} \, \frac{Q}{M} \, \epsilon\Bigr)Mr
 + \Bigl( \frac{q^2}{m^2} - 1 \Bigr) Q^2 \Bigr] \, ,
\e
where $\epsilon = E/m$ is the specific energy of the particle.
Motion is possible only if $\dot{r}^2$ is non-negative. This implies that the radial coordinate of the test
particle must necessarily be outside the region $(r_- , \, r_+)$ where the turning radii $r_\pm$ are given
by \cite{cohen}:
\b
\label{r0}
r_\pm = \frac{M}{\epsilon^2 - 1} \Biggl[ \epsilon \, \frac{q}{m} \, \frac{Q}{M} - 1
 \pm \sqrt{\Bigl( \epsilon \, \frac{q}{m} \, \frac{Q}{M} - 1 \Bigr)^2
- (1 - \epsilon^2) \Bigl(1 - \frac{q^2}{m^2} \Bigr) \frac{Q^2}{M^2}} \, \, \Biggr].
\e
The loci of the event horizon and the Cauchy horizon for the Reissner--Nordstr\" om geometry are:
\b
R_{\pm} = M \biggl( 1 \pm \sqrt{1 - \frac{Q^2}{M^2}} \biggr)\, ,
\e
respectively. For a particle as the electron, $Q/M \sim 10^{21}$. Such solution does not have any horizons
and is called a naked singularity. Naked singularities exhibit gravitational repulsion and this explains the
existence of turning radii. The centre $r = 0$, however, can be reached (see \cite{cohen} for the analysis)
by a suitably charged incoming particle --- when $\mbox{sign}(Q) q/m  \le - 1$. For a positively charged center
for instance $(Q > 0)$, a suitably charged particle (with $q/m \le -1$) will hit the the naked singularity and,
as shown by \cite{cohen}, a naked singularity can be destroyed if sufficient amount of mass and opposite charge
are fed into it. While in existence, this positively charged naked singularity will never be reached by incoming
particles of small negative charge (i.e. $- 1 < q/m < 0$), neutral particles and all positively charged particles
(i.e. $q \ge 0$). For these particles the centre is surrounded by an impenetrable sphere of radius $r_0(T) = r_+$
given by equation (\ref{r0}). The radius of that impenetrable sphere depends on the energy $\epsilon$ of the incoming
particles. For very high energies ($\epsilon \to \infty $), the particle's "radius" can be written as:
\b
\label{r}
r_0(T) = \frac{Q}{\epsilon} \Bigl( \frac{q}{m} + 1 \Bigr) = \frac{Q(q+m)}{kT} \, .
\e
The expansion mechanism is based on the identification of the "unusual" particles with naked singularities. Thus,
the "unusual" particles will have "radii" $r_0(T)$, as determined in equation (\ref{r}). When the temperature starts
to drop, the ``unusual'' particles rapidly increase their ``size'' in inverse proportion with the temperature and
drive apart the ``normal'' fraction of the gas. Thus the Universe will increase its size in inverse proportion with
the temperature: $a(\tau) \sim r_0(\tau) \sim 1/T(\tau) \, ,$ where $a(\tau)$ is the scale factor of the Universe.
We therefore get the usual relation:
\b
aT \,\, = \,\, \mbox{const}
\e
or:
\b
\label{hubble}
H \,\, \equiv \,\, \frac{\dot{a}}{a} \,\, = \,\, \frac{\dot{r}_0}{r_0} \,\, = \,\, - \frac{\dot{T}}{T} \, .
\e
Let us consider the expansion rate equation (see, for example, \cite{peebles}) without cosmological constant:
\b
H^2 = \Bigl(\frac{\dot{a}}{a}\Bigr)^2 = \frac{8 \pi G \rho}{3} + \frac{K}{a^2} \, .
\e
In this equation, $K$ is an integration constant related to the spacetime curvature, while the density $\rho$ includes
all forms. At the present epoch, the main contribution in $\rho$ comes from ordinary matter. In our model, the main
contribution in $\rho$ comes from the electrostatic field of the "unusual" particles. To estimate this contribution, let
us consider a probe which is being driven away by an expanding "unusual" particle. That is, the distance between the
"unusual" particle and the probe is equal to $r_0 (T)$. The intensity of the electric field of the "unusual" particle at
this distance is proportional to $Q/r_0^2(T)$. The energy density of the electric field is proportional to the square of
the intensity of the electric field, that is, the main contribution to $\rho$ comes from a term proportional to
$Q^2/r_0^4(T)$, namely, a term proportional to the fourth power of the temperature (since $r_0(T)$ is inversely
proportional to the temperature). Therefore:
\b
\frac{\dot{a}}{a} \sim T^2 \, .
\e
Substituting (\ref{hubble}) into this equation, one gets:
\b
\frac{\dot{T}}{T} \sim - T^2 \, .
\e
The solution to this equation is $T \sim \sqrt{\frac{1}{\tau}}$ and, therefore, $a(\tau) \sim \sqrt{\tau}$ ---
behaviour corresponding to the expansion during the radiation-dominated era. This is not surprising as the expansion
law was derived for a gas of ultra-relativistic "normal" particles (which, essentially, are pure radiation in view
of the fact that their energies are much higher than their rest masses --- as it is for massless particles), in a
fluid of "unusual" particles. (It would be interesting to study the effects which the "growing" naked singularities
have on the photons.) One should also note that in the derivation of the expansion law $a(\tau) \sim \sqrt{\tau}$, the
charge $Q$ was considered constant (i.e. not changing with time or the temperature). \\
It is very plausible that the "unusual" particles are not stable and that there are various possible annihilation
mechanisms for them. Apart from annihilation into black holes by "capturing" oppositely charged "normal" particles (as
proposed by Cohen and Gautreau in \cite{cohen}),  these superheavy charged particles can annihilate through different
competing mechanisms: they can themselves recombine into neutral particles or decay before or after that. Of particular
importance for such possibilities is the work of Ellis et al. \cite{ellis} on the astrophysical constraints on massive
unstable neutral relic particles and also the work of Gondolo et al. \cite{gondolo} on the constraints of the relic
abundance of a dark matter candidate --- a generic particle of mass in the range of $1 - 10^{14}$ TeV, lifetime greater
than $10^{14} - 10^{18}$ years, decaying into neutrinos. \\
Our assumption however is that the superheavy charged particles have survived annihilation long enough so as to drive the
expansion during the radiation-dominated epoch and, possibly, beyond. To estimate the number density of the "unusual"
particles during the radiation-dominated epoch, let us recall our assumption that the Universe, even though having
globally Robertson--Walker geometry, has local Reissner--Nordstr\"om geometry --- namely, it is a fluid of naked
singularities. It is plausible to assume that these naked singularities are densely packed spheres that fill the entire
Universe and drive its expansion. Thus, the volume $V$ of the Universe, at any moment during the Reissner--Nordstr\"om
expansion, would be of the order of the number $N$ of these particles, times the "volume" of the repulsive sphere of an
"unusual" particle:  $V \sim N r_0^3(T)$. Therefore, the number density of the "unusual"  particles is of the order of
$r_0^{-3}(T)$. \\
Even though our analysis is focused on a purely classical description (in view of the ultra high masses of the "unusual"
particles), one should not be left with the impression that there are no quantum effects in our model. On the
contrary --- quantum effects are present throughout the entire process --- the "normal" fraction is an ideal quantum gas
of massless particles. The laws of quantum theory are present in the neighbourhood of the expanding classical
objects of ultra high mass. Of course, one would also expect quantum interactions between the two fractions and between
the "unusual" particles themselves. Our point however, is that these quantum interactions would not compete with and
prevent the expansion which is due to classical gravitational effects. Study of quantum interactions between all of the
types of particles involved and study of the quantum properties of naked singularities in particular is yet to gain
momentum. This will, undoubtedly, broaden the range of validity of this expansion model. Certain aspects, for example,
of the classical and quantum properties of repulsive singularities have been analyzed in the context of four-dimensional
extended supergravity by Gaida et al. --- see \cite{gaida} and the references therein.

\vskip1cm
\noindent
R.I. Ivanov acknowledges partial support from the Bulgarian National Science Foundation, grant 1410. \\
V.G. Gueorguiev acknowledges partial support under the auspices of the U.S. Department of Energy by the University
of California, Lawrence Livermore National Laboratory under contract No. W-7405-Eng-48.


\begin{thebibliography}{99}


\bibitem{guth} A.N. Guth, in: {\it Carnegie Observatories Astrophysics Series, Volume II: Measuring and Modeling
the Universe}, edited by W.L. Freedman, Cambridge University Press (2002), astro-ph/0404546.

\bibitem{lucchin} F. Lucchin and S. Matarrese, Phys Rev. {\bf D 32 (6)}, 1316--1322 (1985).

\bibitem{brisudova} M. Brisudova, W.H. Kinney, and R. Woodard, Phys. Rev. {\bf D 65}, 103513 (2002), hep-ph/0110174.

\bibitem{rn} H. Reissner, Ann. Phys. ({\it Germany}) {\bf 50}, 106--120 (1916); \newline
G. Nordst\" om, Proc. Kon. Ned. Akad. Wet. {\bf 20}, 1238--1245 (1918).

\bibitem{mtw} C.W. Misner, K.S. Thorne and J. Wheeler, {\it Gravitation}, W.H. Freeman (1973).

\bibitem{cohen} J.M. Cohen and R. Gautreau, Phys. Rev. {\bf D 19 (8)}, 2273--2279 (1979).

\bibitem{rw} H.P. Robertson, Astrophys. J. {\bf 82}, 284--301 (1935); \newline
A.G. Walker, Mon. Not. Roy. Astr. Soc. {\bf 95}, 263--269 (1935).

\bibitem{kerr} R.P. Kerr, Phys. Rev. Lett. {\bf 11}, 237 (1963).

\bibitem{newman} E.T. Newman, R. Couch, K. Chinnapared, A. Exton, A. Prakash, and R. Torrence,
J. Math. Phys. {\bf 6}, 918-919 (1965).

\bibitem{hawk} S. Hawking, Mon. Not. R. Astr. Soc. {\bf 152}, 75--78 (1971).

\bibitem{glashow} A. de Rujula, S.L. Glashow, and U. Sarid, Nucl. Phys. {\bf B 333}, 173 (1990).

\bibitem{preskill} J. Preskill, Phys. Rev. Lett. {\bf 43 (19)}, 1365 (1979).

\bibitem{bl} R.H. Boyer and R.W. Lindquist, J. Math. Phys. {\bf 8 (2)}, 265 (1967).

\bibitem{car} B. Carter, Phys. Rev. {\bf 174}, 1559 (1968).

\bibitem{fn} V. Frolov and I. Novikov, {\it Black Hole Physics: Basic Concepts and New Developments}, Kluwer (1998).

\bibitem{peebles} P.J.E. Peebles, {\it Principles of Physical Cosmology}, Princeton University Press (1993).

\bibitem{ellis} J.R. Ellis, G.B. Gelmini, J.L. Lopez, D.V. Nanopoulos, and S. Sarkar, Nucl. Phys. {\bf B 373}, 399 (1992).

\bibitem{gondolo} P. Gondolo, G.B. Gelmini, and S. Sarkar, Nucl. Phys. {\bf B 392}, 111 (1993), hep-ph/9209236.

\bibitem{gaida} I. Gaida, H.R. Hollmann, and J.M. Stewart, Classical and Quantum Gravity 16, 2231 (1999), hep-th/9811032.

\end{thebibliography}
\end{document}